\documentclass[a4paper,10pt]{article}

\title{A Clifford algebra gauge invariant Lagrangian for gravity. Part 2 : compatibility with General Relativity tests.}
\author{J.P. Pansart\footnote{Retired from : Commissariat \`a l Energie Atomique, CEN Saclay, DSM/Irfu/SPP. Contact : jean-pierre.pansart@orange.fr} \\ \\ Commissariat \`a l’ Energie Atomique, CEN Saclay, DSM/Irfu/SPP \\ 91191  Gif-sur-Yvette, France}

\begin{document}

\maketitle

\section*{Introduction}

  In the last years, there has been a large number of papers discussing possible extensions of the Einstein-Hilbert Lagrangian for gravity : $L = R$ , where $R$ is the scalar space-time curvature. There has been mainly 2 kinds of theories discussed, the so-called $f(R)$ theories, where $f$ is a function of the scalar curvature \cite{SF}, and theories based on quadratic terms built on the curvature tensor \cite{CS}, more or less like the Lagrangian of gauge fields, or even more general ones \cite{BH} including torsion. The existence of torsion has also often been discussed with possible effects for the early universe \cite{Pop}.\\
   In the present note we consider a gauge theory based on the Clifford algebra represented by the Dirac matrices. This leads to a Lagrangian containing four terms :  the Einstein-Hilbert one, a cosmological constant term, a quadratic Lagrangian term and a torsion one. Torsion is naturally introduced because the connexion coefficients and the basis vectors are independent fields.\\
  The goal of this note is to show that there is no contradiction with the experimental observations. The consequences of such an extended Lagrangian would appear only in strong gravitational fields, but we have not yet look at them.\\
  The construction is independent of the number of space-time dimensions. This is discussed in a separate note \cite{JPP}. The reduction from an n-dimensional space-time to the 4-dimensional space-time gives the gravitational Lagrangian described above and the usual quadratic Lagrangian of gauge fields.\\
 The next section sets the notations and recalls very briefly some basic geometrical equations. Section 2 explains the construction of the Lagrangian, and the equations of motion are deduced in section 3. The following sections look at the consequences for, respectively, the static spherical symmetric case, the expanding isotropic universe, and the gravitational field of a rotating body far from it. The appendix A details the consequences of the universe isotropy on the tensor components. The present note, called « part 2 » in the title, can be read independently of the note on higher dimensions, called « part 1 ».

\section{ Notations and basic geometrical equations.}

 The space-time coordinates $\left\{ {{x^\alpha }} \right\}$ of a point $x$ are labelled with Greek letters : $\alpha \,,\,\beta \,,\,\gamma \;...$    , \  $0 \le \alpha \,,\,\beta \,,\,\gamma ,\;...\;\, < n$ . The time coordinate is : ${x^0}$ and, when it is necessary to distinguish spatial coordinates from the time coordinate, the letters : $\mu \,,\,\nu \,,\,\rho \;,\;\eta \;...$ are used.
 The vectors of the local natural frame are written : $\overrightarrow {{e_\alpha }} \;,\;\overrightarrow {{e_\beta }} \;,\;...$ . When tensors are expressed with respect to local orthonormal frames they are labelled with Latin letters : $a\,,\,b\,,\,c\;...$ . The orthonormal local frame basis vectors are called : $\overrightarrow {{h_a}} $ , and we set : $\overrightarrow {{h_a}}  = h_a^\alpha \;\overrightarrow {{e_\alpha }} $ . The metric tensor is $g_{\alpha \beta }$  , and  $g^{\alpha \beta }$  is its inverse.
 The signature of the metric is : $( + \; - \; - \; - )$ . In the case of local orthonormal frames,  the metric tensor is written : ${\eta _{a\,b}}$  and its diagonal terms are : ${\eta _{a a}} = ( + 1\,,\; - 1\,,\; - 1\,,\; - 1)$ . The commutator of the vectors $\left\{ {\overrightarrow {{h_a}} } \right\}$ is :
${\left[ {{h_a}\,,\;{h_b}} \right]^\gamma } = h_a^\alpha \,{\partial _\alpha }h_b^\gamma  - h_b^\alpha \,{\partial _\alpha }h_a^\gamma  = C_{\,.\;a{\kern 1pt} b}^c\,h_c^\gamma$
\hspace{40pt} (1.1)\\

 In the neighborhood of a given point, the local coordinates, with respect to the local orthonormal frame attached to this point, are given by the 1-forms : ${\omega ^a} = h_\alpha ^a\;d{x^\alpha }$ , which satisfy the structure equations : \\
\hspace*{100pt}
  $d{\omega ^a} + \omega _{\,.\;b}^a \wedge {\omega ^b} = {\Sigma ^a}$
\hspace{105pt} (1.2)\\                                  
where : $\omega _{\,.\;b}^a = \omega _{\,.\;b{\kern 1pt} \gamma }^a\;d{x^\gamma }$ are the connexion 1-forms and ${\Sigma ^a}$ is the torsion 2-form. We shall also write: 
$\omega _{\,.\;b}^a = \omega _{\,.\;b{\kern 1pt} c}^a\,{\omega ^c}\;\; \leftrightarrow \;\;\omega _{\,.\;b{\kern 1pt} c}^a = \omega _{\,.\;b{\kern 1pt} \gamma }^a\,h_c^\gamma $ . The connexion 1-forms are related to the connexion coefficients by :\\ 
\hspace*{80pt}
$\omega _{\,.\;b{\kern 1pt} \gamma }^a = \Gamma _{\,.\;\beta {\kern 1pt} \gamma }^\alpha \,h_\alpha ^a\,h_b^\beta  + \,h_\delta ^a\,{\partial _\gamma }h_b^\delta$ 
\hspace{85pt} (1.3)\\

 The connexion coefficients are the sum of two terms :\\
\hspace*{100pt} $\Gamma _{\,.\;\beta {\kern 1pt} \gamma }^\alpha  = \widetilde \Gamma _{\,.\;\beta {\kern 1pt} \gamma }^\alpha  + \overline S _{\,.\;\beta {\kern 1pt} \gamma }^\alpha $
\hspace{100pt} (1.4)\\                                                                                                                        
where the first term is the Christoffel symbol and the second is the contorsion tensor. The contorsion is anti symmetric with respect to the two first indices : $\overline S {_{\alpha \,\beta \,\gamma }} + \overline S {_{\beta \,\alpha \,\gamma }} = 0$ . The torsion tensor is : \\
\hspace*{60pt}  $S_{{\kern 1pt} .{\kern 1pt} \beta {\kern 1pt} \gamma }^\alpha  = \frac{1}{2}\;(\Gamma _{{\kern 1pt} .{\kern 1pt} \beta {\kern 1pt} \gamma }^\alpha  - \Gamma _{{\kern 1pt} .{\kern 1pt} \gamma {\kern 1pt} \beta }^\alpha ) = \frac{1}{2}\;(\overline S _{{\kern 1pt} .{\kern 1pt} \beta {\kern 1pt} \gamma }^\alpha  - \overline S _{{\kern 1pt} .{\kern 1pt} \gamma {\kern 1pt} \beta }^\alpha )$
\hspace{42pt} (1.5a)\\                                              
 and inversely :
$\overline S _{{\kern 1pt} .{\kern 1pt} \beta {\kern 1pt} \gamma }^\alpha  = S_{{\kern 1pt} .{\kern 1pt} \beta {\kern 1pt} \gamma }^\alpha  - S_{{\kern 1pt} \beta \,.{\kern 1pt} {\kern 1pt} \gamma }^{\;\;\,\alpha } - S_{{\kern 1pt} \gamma \,.{\kern 1pt} {\kern 1pt} \beta }^{\;\;\,\alpha }\quad ;\quad S_{{\kern 1pt} \beta \,.{\kern 1pt} {\kern 1pt} \gamma }^{\;\;\,\alpha } = {g^{\alpha \,\delta }}\,{S_{\beta \,\delta \,\gamma }}$
\hspace{5pt} (1.5b)\\
The torsion 2-form is :
${\Sigma ^a} = \Sigma _{\,.\;b{\kern 1pt} c}^a\,{\omega ^b} \wedge {\omega ^c} =  - h_\alpha ^a\,S_{\,.\;\beta {\kern 1pt} \gamma }^\alpha \,d{x^\beta } \wedge d{x^\gamma }$
\hspace{15pt} (1.6)\\
Using (1.3) we set :
\hspace{15pt}
$\widetilde \Gamma _{\,.\;b{\kern 1pt} \gamma }^a = \widetilde \Gamma _{\,.\;\beta {\kern 1pt} \gamma }^\alpha \,h_\alpha ^a\,h_b^\beta  + \,h_\delta ^a\,{\partial _\gamma }h_b^\delta $
\hspace{60pt} (1.7)\\ 
The curvature 2-form is defined by :\\
\hspace*{55pt}
$\Omega _{\,.\;b}^a = d\omega _{\,.\;b}^a + \omega _{\,.\;c}^a \wedge \omega _{\,.\;b}^c = R_{\,\,.\;\,b\;c\,d}^a\;{\omega ^c} \wedge {\omega ^d}$
\hspace{55pt} (1.8)\\


\section{ The gravitational field as a gauge field.}

 Let us consider the Dirac matrices ${\gamma ^a}$ which represent the basis elements $\{ {e_a}\} $ of the Clifford algebra : \hspace{10pt}  ${e_a}{\kern 1pt} .{\kern 1pt} {\kern 1pt} {e_b} + {e_b}{\kern 1pt} .{\kern 1pt} {\kern 1pt} {e_a} = 2\,{\eta _{a{\kern 1pt} b}}$\\                
The commutators :  ${R^{a{\kern 1pt} b}} = {\textstyle{1 \over 4}}\,\left[ {{\gamma ^a},\;{\gamma ^b}} \right]$  represent the generators of the rotation group. They satisfy :\\
\hspace*{30pt}
$[{R^{a{\kern 1pt} b}},\,{R^{c{\kern 1pt} d}}] =  - {\eta ^{a{\kern 1pt} d}}{R^{c{\kern 1pt} b}} - {\eta ^{a{\kern 1pt} c}}{R^{b{\kern 1pt} d}} - {\eta ^{b{\kern 1pt} d}}{R^{a{\kern 1pt} c}} - {\eta ^{b{\kern 1pt} c}}{R^{d{\kern 1pt} a}}$
\hspace{30pt} (2.1a)\\
\hspace*{107pt}
$[{\gamma ^a},{\gamma ^b}] = 4\,{R^{a{\kern 1pt} b}}$
\hspace*{115pt} (2.1b)\\
\hspace*{82pt}
$[{\gamma ^a},\,{R^{c{\kern 1pt} d}}] = {\eta ^{a{\kern 1pt} c}}{\gamma ^d} - {\eta ^{a{\kern 1pt} d}}{\gamma ^c}$
\hspace{90pt} (2.1c)\\
This set of relations is a graded Lie algebra which satisfies Jacobi’s identities.  The ${\gamma ^a}$ and ${R^{a{\kern 1pt} b}}$ matrices are the elements of  a representation $\Gamma $ of an algebra defined by the relations (2.1) and named $\Gamma ({X_x})$ , where ${X_x}$ are the basis elements of this algebra. To this algebra we associate a gauge field: \\ 
\hspace*{90pt}
$W = \alpha \,{\omega _{a{\kern 1pt} b}}\,{R^{a{\kern 1pt} b}} + \beta \,{\omega _a}\,{\gamma ^a}$
\hspace{90pt} (2.2)\\
where : ${\omega _{a{\kern 1pt} b}} =  - {\omega _{b{\kern 1pt} a}}$ and ${\omega _a}$ are differential forms of degree 1 :\\
\hspace*{50pt}
${\omega _{a{\kern 1pt} b}} = {\omega _{a{\kern 1pt} b{\kern 1pt} \alpha }}\,d{x^\alpha }$
\hspace{25pt}
${\omega _a} = {\eta _{a{\kern 1pt} b}}\,{\omega ^b} = {\eta _{a{\kern 1pt} b}}\,h_\alpha ^b\,d{x^\alpha }$
\hspace{30pt} (2.3)\\
and : $\alpha$ , $\beta$  are arbitrary constants.\\
 The meaning of these gauge fields is the following :  the fields ${\omega _{a{\kern 1pt} b}}$  are the connexion coefficients defined with respect to a family of  local orthonormal frames, and the 1-forms
${\omega ^a}(x) = {\eta ^{a{\kern 1pt} b}}\,{\omega _b}(x)$ 
  are the coordinates, in the neighborhood of a given point $x$ , with respect to these frames. This can be understood by computing the curvature 2-form (1.8) :\\
\hspace*{110pt}
 $G = dW + W \wedge W$
\hspace{100pt} (2.4)\\
$G = (\alpha \,d{\omega _{a{\kern 1pt} b}}\,{R^{a{\kern 1pt} b}} + \beta \,d{\omega _a}\,{\gamma ^a}) + (\alpha \,{\omega _{c{\kern 1pt} d}}\,{R^{c{\kern 1pt} d}} + \beta \,{\omega _c}\,{\gamma ^c}) \wedge (\alpha \,{\omega _{e{\kern 1pt} f}}\,{R^{e{\kern 1pt} f}} + \beta \,{\omega _e}\,{\gamma ^e})$ \\
\hspace*{5pt}
$G = \alpha \,d{\omega _{a{\kern 1pt} b}}\,{R^{a{\kern 1pt} b}} + \frac{{{\alpha ^2}}}{2}\;({\omega _{c{\kern 1pt} d}} \wedge {\omega _{e{\kern 1pt} f}}\,{R^{c{\kern 1pt} d}}\,{R^{e{\kern 1pt} f}} + {\omega _{e{\kern 1pt} f}} \wedge {\omega _{c{\kern 1pt} d}}\,{R^{e{\kern 1pt} f}}\,{R^{c{\kern 1pt} d}})$
\\
\hspace*{15pt}
 $\quad \quad \quad  + \beta \,d{\omega _a}\,{\gamma ^a} + \alpha {\kern 1pt} \beta \;({\omega _{c{\kern 1pt} d}} \wedge \,{\omega _e}\,{R^{c{\kern 1pt} d}}\,{\gamma ^e} + \,{\omega _e} \wedge {\omega _{c{\kern 1pt} d}} \wedge \,{\gamma ^e}{R^{c{\kern 1pt} d}}) + {\beta ^2}\,{\omega _c} \wedge \,{\omega _e}\;{\gamma ^c}\,{\gamma ^e}$
\\
and since the gauge fields (2.3) are 1-forms, one has :\\
\hspace*{5pt}
$G = \alpha \,d{\omega _{a{\kern 1pt} b}}\,{R^{a{\kern 1pt} b}} + \frac{{{\alpha ^2}}}{2}\;{\omega _{c{\kern 1pt} d}} \wedge {\omega _{e{\kern 1pt} f}}\,[{R^{c{\kern 1pt} d}},\,{R^{e{\kern 1pt} f}}] + \beta \,d{\omega _a}\,{\gamma ^a}$ \\
\hspace*{60pt}
$ + \alpha {\kern 1pt} \beta \;{\omega _{c{\kern 1pt} d}} \wedge \,{\omega _e}\,[{R^{c{\kern 1pt} d}},\,{\gamma ^e}] + 2{\beta ^2}\,{\omega _a} \wedge \,{\omega _b}\,{R^{a{\kern 1pt} b}}$\\
With the relations (2.1) this becomes :\\
\hspace*{20pt}
$G = [\alpha \,d{\omega _{a{\kern 1pt} b}} + 2\,{\alpha ^2}{\kern 1pt} {\omega _{a{\kern 1pt} f}} \wedge \omega _{\,.\,b}^f + 2{\beta ^2}\,{\omega _a} \wedge \,{\omega _b}]\,{R^{a{\kern 1pt} b}}$ \\
\hspace*{110pt}
$ + \beta \,\,[d{\omega _a} + 2\,\alpha \;\omega _{a\,.}^{\;e} \wedge \,{\omega _e}]\;{\gamma ^a}$
\hspace{60pt} (2.5)\\
We can set  :
\hspace{80pt}
$\alpha  = \frac{1}{2}$
\hspace{125pt} (2.6)\\
Then :
\hspace{5pt}
$G = \frac{1}{2}\;[d{\omega _{a{\kern 1pt} b}} + \;{\kern 1pt} {\omega _{a{\kern 1pt} f}} \wedge \omega _{\,.\,b}^f + 4{\beta ^2}\,{\omega _a} \wedge \,{\omega _b}]\,{R^{a{\kern 1pt} b}}$ \\
\hspace*{112pt}
$ + \beta \,\,[d{\omega _a} + \;\omega _{a\,.}^{\;e} \wedge \,{\omega _e}]\;{\gamma ^a}$
\hspace{70pt} (2.7a)\\
 With the above interpretation of the gauge fields, the first two terms in the first brackets correspond to the usual curvature 2-form : $\Omega _{\,.\;b}^a = d\omega _{\,.\;b}^a + \;{\kern 1pt} \omega _{\,.\;c}^a \wedge \omega _{\,.\;b}^c$ , and the second brackets contains the structure equations :  $d{\omega ^a} + \;\omega _{\,.\;b}^a \wedge \,{\omega ^b} = {\Sigma ^a}$ , where  ${\Sigma ^a}$ represents the torsion 2-form (1.6) : \\ 
\hspace*{55pt}
$G = \frac{1}{2}\;[{\Omega _{a{\kern 1pt} b}} + 4{\beta ^2}\,{\omega _a} \wedge \,{\omega _b}]\,{R^{a{\kern 1pt} b}} + \beta \,\,{\Sigma _a}\;{\gamma ^a}$
\hspace{50pt} (2.7b)\\
\\
The standard minimum gauge field Lagrangian is :\\
\hspace*{110pt}
$L = Tr(G \wedge  * \,G)$
\hspace{105pt} (2.8)\\
where : $ * $ is the Hodge’s star operator. Taking into account the ${\gamma ^a}$ matrices properties : 
\hspace{40pt}
$Tr({R^{a{\kern 1pt} b}}\,{R^{c{\kern 1pt} d}}) = \frac{N}{4}({\eta ^{a{\kern 1pt} d}}\,{\eta ^{b{\kern 1pt} c}} - {\eta ^{a{\kern 1pt} c}}\,{\eta ^{b{\kern 1pt} d}})$ \\
\hspace*{75pt}
$Tr({R^{a{\kern 1pt} b}}\,{\gamma ^c}) = 0$
\hspace{30pt}
$Tr({\gamma ^a}\,{\gamma ^b}) = N\,{\eta ^{a{\kern 1pt} b}}$
\hspace{30pt} (2.9)\\
where : $N$ is the spinor dimension,  one has :\\
$L/N =  - \;{\beta ^2}\;{\Omega ^{a{\kern 1pt} b}} \wedge  * \,({\omega _a} \wedge \,{\omega _b})\; - \;\frac{1}{8}\;{\Omega ^{a{\kern 1pt} b}} \wedge  * \,{\Omega _{a{\kern 1pt} b}}$
\hspace{75pt}  (2.10a) \\
$- 2\,{\beta ^4}\,({\omega ^a} \wedge \,{\omega ^b}) \wedge  * \,({\omega _a} \wedge \,{\omega _b}) + {\kern 1pt} {\beta ^2}\,\,{\eta _{a\,b}}\,(d{\omega ^a} + \;\omega _{\,.\;c}^a \wedge \,{\omega ^c}) \wedge  * \,(d{\omega ^b} + \;\omega _{\,.\;d}^b \wedge \,{\omega ^d})$
\\
  The first term correspond to the Einstein-Hilbert Lagrangian of General Relativity. The third term represents the contribution of a cosmological constant, since this term is proportional to the volume element. The second term is quadratic and has the form of  standard gauge field Lagrangian. The equation (2.10a) can also be re-written :\\
\hspace*{10pt}
$L/N =  - \;{\beta ^2}\;{\Omega ^{a{\kern 1pt} b}} \wedge  * \,({\omega _a} \wedge \,{\omega _b})\; - \;\frac{1}{8}\;{\Omega ^{a{\kern 1pt} b}} \wedge  * \,{\Omega _{a{\kern 1pt} b}}$
\hspace{58pt} (2.10b)\\
\hspace*{5pt}
$ - 2\,{\beta ^4}\,({\omega ^a} \wedge \,{\omega ^b}) \wedge  * \,({\omega _a} \wedge \,{\omega _b}) + {\kern 1pt} {\beta ^2}\,\,{\Sigma ^a} \wedge  * \,{\Sigma _a} + \mu \;(d{\omega ^a} + \;\omega _{\,.\;b}^a \wedge \,{\omega ^b} - {\Sigma ^a})$  \\
where : $\mu$ is a Lagrange multiplicator. \\
In (2.10) the torsion is introduced naturally, not as an extra field, this is a direct consequence of the definition (2.2) where ${\omega _a}$ and ${\omega _{a{\kern 1pt} b}}$ are independent fields. The equations of motion are calculated in the next chapter.\\

  Until now, we have discussed the interpretation of the gauge fields introduced in (2.2), but it remains to see how these fields transform. Taking into account the algebra (2.1) , one considers the infinitesimal  transformations :  $S = I + i{\kern 1pt} {\varepsilon _a}{\kern 1pt} {\gamma ^a} + {\varepsilon _{a{\kern 1pt} b}}\,{R^{a{\kern 1pt} b}}$   , where : ${\varepsilon _{a{\kern 1pt} b}} =  - {\varepsilon _{b{\kern 1pt} a}}$ and : $\left| {{\varepsilon _a}} \right|{\kern 1pt} ,\,\;\left| {{\varepsilon _{a{\kern 1pt} b}}} \right|\, \ll 1$  . The transformation law of gauge field is :  $W' = {S^{ - 1}}{\kern 1pt} W\,S + {S^{ - 1}}dS$  , which gives : $G' = {S^{ - 1}}{\kern 1pt} G\,S$  and makes the Lagrangian (2.8) invariant. With, at first order : ${S^{ - 1}} = I - i{\kern 1pt} {\varepsilon _a}{\kern 1pt} {\gamma ^a} - {\varepsilon _{a{\kern 1pt} b}}\,{R^{a{\kern 1pt} b}}$ , one has, still at first order :\\
\hspace*{5pt}
$W' = W + i\,\alpha \,{\omega _{a{\kern 1pt} b}}{\kern 1pt} {\varepsilon _e}\;\left[ {{R^{a{\kern 1pt} b}},\;{\gamma ^e}} \right] + \alpha \;{\omega _{a{\kern 1pt} b}}{\kern 1pt} {\varepsilon _{ef}}\;\left[ {{R^{a{\kern 1pt} b}},\;{R^{e{\kern 1pt} f}}} \right] + i\,\beta {\kern 1pt} {\kern 1pt} {\omega _a}{\kern 1pt} {\varepsilon _e}\;\left[ {{\gamma ^a},{\gamma ^e}} \right]$
\\
\hspace*{10pt}
$ + \;\beta \;{\omega _a}{\kern 1pt} {\varepsilon _{ef}}\;\left[ {{\gamma ^a},\;{R^{e{\kern 1pt} f}}} \right] + i\,d{\varepsilon _e}\,{\gamma ^e} + d{\varepsilon _{ef}}\;{R^{e{\kern 1pt} f}}$
\\
then , with the algebra (2.1) :\\
\hspace*{30pt}
$\omega {'_a} = \;{\omega _a}{\kern 1pt}  + {\varepsilon _{e{\kern 1pt} a}}{\kern 1pt} {\omega ^e}\; - {\varepsilon _{a{\kern 1pt} e}}{\kern 1pt} {\omega ^e} + \frac{i}{\beta }\,(d{\varepsilon _a} + \alpha \,\omega _{a\,.}^{\;\,e}\,{\varepsilon _e} - \alpha \,\omega _{\;.\;a}^e\,{\varepsilon _e})$
\\
and with (2.6) : \hspace{20pt} $\omega {'_a} = \;{\omega _a}{\kern 1pt}  + {\varepsilon _{e{\kern 1pt} a}}{\kern 1pt} {\omega ^e}\; - {\varepsilon _{a{\kern 1pt} e}}{\kern 1pt} {\omega ^e} + \frac{i}{\beta }\,D{\varepsilon _a}$
\\
If : $D{\varepsilon _a} = 0$ , ${\omega ^a}$ transforms like a vector under an (infinitesimal) rotation whose coefficients are the ${\varepsilon _{e{\kern 1pt} f}}$ . In that case : ${\varepsilon _a} = 0$  and the field ${\omega _{a{\kern 1pt} b}}$ transforms like a gauge field with respect to rotations. The restriction of the gauge transformation makes ${\omega ^a}$ transform like a vector.  The 1-forms : ${\omega ^a} = h_\alpha ^a\,d{x^\alpha }$ are directly related to the basis vectors of the local frames.\\

In the above description ${\omega ^a}$ and ${\omega _{a{\kern 1pt} b}}$ are independent gauge fields. How does that modifies the Lagrangian of matter fields ? We shall suppose that ordinary matter fields are spinor fields, and therefore we shall consider the covariant derivative of such fields.\\
Let $\psi $ be a spinor field. The covariant derivative of a spinor field with gauge fields is :\\
\hspace*{20pt}
$D\psi  = d\psi  + {\textstyle{1 \over 4}}\,{\omega _{c{\kern 1pt} d}}\,{\gamma ^c}{\kern 1pt} {\gamma ^d}\,\psi  + \beta \,{\omega _a}\,{\gamma ^a}\,\psi $ \\
\hspace*{50pt}
$ \to \quad {D_{{\kern 1pt} \alpha }}\psi  = {\partial _\alpha }\psi  + {\textstyle{1 \over 4}}\,{\omega _{c{\kern 1pt} d{\kern 1pt} \alpha }}\,{\gamma ^c}{\kern 1pt} {\gamma ^d}\,\psi  + \beta \,h_\alpha ^b\,{\eta _{b{\kern 1pt} c}}\,{\gamma ^c}\,\psi $ \\
The last term is not present in the usual covariant derivative of a spinor field. The Lagrangian of such a field is :    
\hspace{10pt}
$L = \overline {\psi \,} h_a^\alpha \,{\gamma ^a}\,i\,{D_\alpha }\psi  + \;h.\,c.$ \\               
where : $h.\,c.$ means : Hermitic conjugate. If  one uses the above covariant derivative, the contribution of the unwanted terms is (if $\beta$ is real) :\\
$i\,\beta \,\overline {\psi \,} h_a^\alpha \,{\gamma ^a}h_\alpha ^b\,{\eta _{b{\kern 1pt} c}}\,{\gamma ^c}\,\psi  + \;h.\,c. = i\,\beta \,\overline {\psi \,} \,{\gamma ^a}\,{\eta _{a{\kern 1pt} c}}\,{\gamma ^c}\,\psi  + \;h.\,c. \sim \,i\,\beta \,\overline {\psi \,} \,\psi  + \;h.\,c. = 0$ \\
Therefore, the gauge field (2.2) gives the usual spinor field Lagrangian built with the usual covariant derivative :\hspace{10pt}
${D_\alpha }\psi  = {\partial _\alpha }\psi  + {\textstyle{1 \over 4}}\,{\omega _{c{\kern 1pt} d{\kern 1pt} \alpha }}\,{\gamma ^c}{\kern 1pt} {\gamma ^d}\,\psi $ \\

Remark : the above Lagrangian of a spinor field is built using the generator ${\gamma ^a}$ of the algebra (2.1). Why not use the generators of type ${R^{a{\kern 1pt} b}}$ instead ? A possible Lagrangian could be :  $L = \overline {{D_a}\psi \,} \,{R^{a{\kern 1pt} b}}\,{D_b}\psi  + \;h.\,c.$    , which gives the second order Dirac equation. In conclusion, the gauge field (2.2) does not introduces unwanted terms.

  The above calculations are independent of the space-time dimension which hereafter is 4. We show, in another note \cite{JPP}, that, following the Kaluza-Klein program, one can recover both the quadratic Lagrangian of gauge fields and the Lagrangian (2.10) for the gravitation.\\

Summary of this section.\\
 The gauge field (2.2) associated to the algebra (2.1) leads to a  gravitational Lagrangian which is more general than the Einstein-Hilbert one, introducing naturally a quadratic term and a cosmological constant. The 1-form fields ${\omega ^a}$ and ${\omega _{a{\kern 1pt} b}}$ are independent of each other, and as a consequence, torsion may exist as an independent field.


\section{The Equations of motion.}

  We now re write the Lagrangian (2.10) in a slightly more general form which puts the Einstein-Hilbert Lagrangian as the main term : \\
\hspace*{2pt}
$ - \,\frac{L}{{{\beta ^2}\,N}} = \;\;{\Omega ^{a{\kern 1pt} b}} \wedge  * \,({\omega _a} \wedge \,{\omega _b})\; + \;\eta \;{\Omega ^{a{\kern 1pt} b}} \wedge  * \,{\Omega _{a{\kern 1pt} b}} + \lambda \;dV + \mu \,\,{\Sigma ^a} \wedge  * \,{\Sigma _a}$
\hspace{5pt} (3.1)\\
with the constraint :   ${\Sigma ^a} = d{\omega ^a} + \;\omega _{\,.\;b}^a \wedge \,{\omega ^b}$  ,  and where : $dV = {\omega _a} \wedge  * {\omega ^a}$ .\\
The parameters $\eta \,,\;\lambda \,,\;\mu $ are now free, although they should be linked by :\\
\hspace*{50pt}
$\eta  = 1/(8\,{\beta ^2})$  \hspace{30pt}  $\lambda  = 2\,{\beta ^2}$
\hspace{30pt} $\mu=-1$
\hspace{55pt} (3.2)\\
  The first term of (3.1) is the Einstein-Hilbert Lagrangian, it is : ${L_{EH}} = R\,dV$ . The second term is : $\eta \,{L_Q} = \eta \;{\Omega ^{a{\kern 1pt} b}} \wedge  * \,{\Omega _{a{\kern 1pt} b}} = \eta \;R_{\;.\,\;.\,\;c{\kern 1pt} d}^{a{\kern 1pt} b}\,R_{a{\kern 1pt} b\;\,.\;\,.}^{\,.\;\,.\;c{\kern 1pt} d}$  \\
The variation of the first term is :\\
\hspace*{30pt}
$\delta {L_{EH}} =  - \delta {\omega ^{a{\kern 1pt} b}} \wedge {\varepsilon _{a{\kern 1pt} b{\kern 1pt} c{\kern 1pt} d}}\,\,{\omega ^c} \wedge (\Sigma _{\;e{\kern 1pt} f}^d{\kern 1pt} {\omega ^e} \wedge {\omega ^f})$  \hspace{80pt}  (3.3) \\
\hspace*{40pt}
$\quad \quad \; + {\Omega ^{a{\kern 1pt} b}} \wedge {\varepsilon _{a{\kern 1pt} b{\kern 1pt} c{\kern 1pt} d}}\,{\omega ^c} \wedge \delta {\omega ^d} + d({\textstyle{1 \over 2}}\;\delta {\omega ^{a{\kern 1pt} b}} \wedge {\varepsilon _{a{\kern 1pt} b{\kern 1pt} c{\kern 1pt} d}}\,{\omega ^c} \wedge {\omega ^d})$\\
 The variation of the quadratic term is : \\
\hspace*{60pt}
$\delta {L_Q} = 2\,\delta {\omega _{a{\kern 1pt} b}} \wedge D * {\Omega ^{a{\kern 1pt} b}} + 2\,d(\delta {\omega _{a{\kern 1pt} b}} \wedge  * {\Omega ^{a{\kern 1pt} b}})$
\hspace{52pt}  (3.4) \\
where : $\Omega _{{\kern 1pt} .\;b}^a = d\omega _{{\kern 1pt} .\;b}^a + \omega _{{\kern 1pt} .\;c}^a \wedge \omega _{{\kern 1pt} .\;b}^c = R_{\,\,.\;\,.\;c\,d}^{a{\kern 1pt} b}\;{\omega ^c} \wedge {\omega ^d}$ is the curvature 2-form. This equation can be transformed into :\\
$\delta {L_Q} = \;2\,d(\delta {\omega _{a{\kern 1pt} b}} \wedge  * {\Omega ^{a{\kern 1pt} b}}) + $
\hspace{185pt} (3.5)\\
 $4\;\delta {\omega _{a{\kern 1pt} b}} \wedge \left\{ {{\partial _d}{R^{a{\kern 1pt} b{\kern 1pt} c{\kern 1pt} d}} + \Gamma _{\,.\;e{\kern 1pt} d}^a\,{R^{e{\kern 1pt} b{\kern 1pt} c{\kern 1pt} d}} + \Gamma _{\,.\;e{\kern 1pt} d}^b\,{R^{a{\kern 1pt} e{\kern 1pt} c{\kern 1pt} d}} + \widetilde \Gamma _{\,.\;e{\kern 1pt} d}^c\,{R^{a{\kern 1pt} b{\kern 1pt} e{\kern 1pt} d}} + \widetilde \Gamma _{\,.\;e{\kern 1pt} d}^d\,{R^{a{\kern 1pt} b{\kern 1pt} c{\kern 1pt} e}}} \right\} \wedge  * {\omega _c}$\\

The variation of the last term of equation (3.1) : ${L_\Sigma } = {\Sigma ^a} \wedge  * \,{\Sigma _a}$  is :\\
\hspace*{60pt}
$\delta {L_\Sigma } = 2\;\left( {d\delta {\omega ^a} + \;\delta \omega _{\,.\;c}^a \wedge \,{\omega ^c} + \omega _{\,.\;c}^a \wedge \,\delta {\omega ^c}} \right) \wedge  * {\Sigma _a}$\\
then, the variation with respect to the connexion is :\\
${\delta _\Gamma }{L_\Sigma } = 2\;\delta {\omega _{a\,c}} \wedge \,{\omega ^c} \wedge  * \Sigma _{\,.\;e{\kern 1pt} f}^a\,{\omega ^e} \wedge {\omega ^f} = 2\;\delta {\omega _{a\,c\,\gamma }}\;h_c^\gamma \,(\overline S {^{a{\kern 1pt} c{\kern 1pt} b}} - \,\overline S {^{a{\kern 1pt} b{\kern 1pt} c}})\;dV$
\hspace{2pt} (3.6)\\
and the variation with respect to the ${\omega ^a}$ is :\\
\hspace*{60pt} 
${\delta _h}{L_\Sigma } = 2\;\delta {\omega ^a} \wedge D( * {\Sigma _a}) + 2\;d(\delta {\omega ^a} \wedge  * {\Sigma _a})$
\hspace{50pt} (3.7)\\
where : $D( * {\Sigma _a}) = d( * {\Sigma _a}) - \omega _{\,.\;a}^b \wedge  * {\Sigma _b}$\\
After some manipulations, one obtains :\\
${\delta _h}{L_\Sigma } = 8\;\delta h_\alpha ^a\,h_g^\alpha \;\left[ {{D_c}\Sigma _{a\;.\;\,.}^{\;\;g{\kern 1pt} c} + 2\,\Sigma _{a\;.\;\,.}^{\;\;g{\kern 1pt} c}\;\Sigma _{\,.\;c{\kern 1pt} d}^d + \Sigma _{a\;.\;\,.}^{\;\;e{\kern 1pt} c}\,\Sigma _{\,.\;e{\kern 1pt} c}^g} \right]\;dV $\\
\hspace*{120pt}
$+ exact\;diff.$
\hspace{112pt} (3.8)\\

 The next task is to transform these variational equations into equations involving tensors written with their components. For that purpose we use (1.3).
The connexion coefficients are (1.4) :  $\Gamma _{\,.\;\beta {\kern 1pt} \gamma }^\alpha  = \widetilde \Gamma _{\,.\;\beta {\kern 1pt} \gamma }^\alpha  + \overline S _{\,.\;\beta {\kern 1pt} \gamma }^\alpha $ where the first term is the Christoffel symbol and the second is the contorsion tensor, therefore :\\
\hspace*{60pt}
$\omega _{\,.\;b{\kern 1pt} \gamma }^a = \,h_\delta ^a\,{\widetilde D_\gamma }h_b^\delta  + \overline S _{\,.\;\beta {\kern 1pt} \gamma }^\alpha \,h_\alpha ^a\,h_b^\beta  = h_\delta ^a\,{\widetilde D_\gamma }h_b^\delta  + \overline S _{\,.\;b{\kern 1pt} \gamma }^a$\\
where : $\widetilde D$ is the covariant derivative built with the Christoffel symbol only.\\
 The covariant derivative of a tensor is :\\
$D\,{T^{{\alpha _1}\;...\;{\alpha _p}}} = d\,{T^{{\alpha _1}\;...\;{\alpha _p}}} + \Gamma _{\,.\;{\beta _i}\,\gamma }^{{\alpha _i}}\,d{x^\gamma }\;{T^{{\alpha _1}\;...\;{\beta _i}\;...\;{\alpha _p}}}$\\
\hspace*{100pt}
$ = \widetilde D\,{T^{{\alpha _1}\;...\;{\alpha _p}}} + \overline S _{\,.\;{\beta _i}\,\gamma }^{{\alpha _i}}\,d{x^\gamma }\;{T^{{\alpha _1}\;...\;{\beta _i}\;...\;{\alpha _p}}}$\\
Since $D\,{T^{{\alpha _1}\;...\;{\alpha _p}}}$ and the last term are tensors, $\widetilde D\,{T^{{\alpha _1}\;...\;{\alpha _p}}}$ is also a tensor. This allows to express the various quantities indifferently in the orthonormal local frames or with coordinate indices.\\
  In the following, we separate the terms involving the Christoffel symbols only from those involving the contorsion.  Please note that, the variable used is the contorsion tensor, not the torsion. First, the curvature tensor is decomposed as :\\
$2\,{R_{a{\kern 1pt} b{\kern 1pt} f{\kern 1pt} g}} = h_f^\alpha \,{\partial _\alpha }{\widetilde \Gamma _{a{\kern 1pt} b{\kern 1pt} g}} - h_g^\alpha \,{\partial _\alpha }{\widetilde \Gamma _{a{\kern 1pt} b{\kern 1pt} f}} + {\widetilde \Gamma _{a{\kern 1pt} e{\kern 1pt} f}}\,\widetilde \Gamma _{\,.\;b{\kern 1pt} g}^e - {\widetilde \Gamma _{a{\kern 1pt} e{\kern 1pt} g}}\,\widetilde \Gamma _{\,.\;b{\kern 1pt} f}^e - {\widetilde \Gamma _{a{\kern 1pt} b{\kern 1pt} e}}\,C_{\,.\;f{\kern 1pt} g}^e$\\
\hspace*{60pt}
$ + \;\;h_f^\alpha \,{D_\alpha }\overline S {_{a{\kern 1pt} b{\kern 1pt} g}} - h_g^\alpha \,{D_\alpha }\overline S {_{a{\kern 1pt} b{\kern 1pt} f}} + \overline S {_{a{\kern 1pt} e{\kern 1pt} f}}\,\overline S _{\,.\;b{\kern 1pt} g}^e - \overline S {_{a{\kern 1pt} e{\kern 1pt} g}}\,\overline S _{\,.\;b{\kern 1pt} f}^e$\\
where the commutation coefficients $C_{\,.\;f{\kern 1pt} g}^e$ have been defined in (1.1) . This relation is simply :\\ 
$2\,{R_{a{\kern 1pt} b{\kern 1pt} f{\kern 1pt} g}} = h_a^\alpha \,h_b^\beta \,h_f^\varphi \,h_g^\gamma \,$\\
\hspace*{30pt}
$\left[ {2\,{{\widetilde R}_{\alpha {\kern 1pt} \beta {\kern 1pt} \varphi {\kern 1pt} \gamma }} + \;\;{{\widetilde D}_\varphi }\overline S {_{\alpha {\kern 1pt} \beta {\kern 1pt} \gamma }} - {{\widetilde D}_\gamma }\overline S {_{\alpha {\kern 1pt} \beta {\kern 1pt} \varphi }}
+ \overline S {_{\alpha {\kern 1pt} \delta {\kern 1pt} \varphi }}\,\overline S _{\,.\;\beta {\kern 1pt} \gamma }^\delta  - \overline S {_{\alpha {\kern 1pt} \delta {\kern 1pt} \gamma }}\,\overline S _{\,.\;\beta {\kern 1pt} \varphi }^\delta } \right]$\\
where : $\widetilde D$ is, as above, the covariant derivative involving Christoffel symbols only. We set :\\
\hspace*{80pt}
${R_{\alpha {\kern 1pt} \beta {\kern 1pt} \varphi {\kern 1pt} \gamma }} = {\widetilde R_{\alpha {\kern 1pt} \beta {\kern 1pt} \varphi {\kern 1pt} \gamma }} + {K_{\alpha {\kern 1pt} \beta {\kern 1pt} \varphi {\kern 1pt} \gamma }}$
\hspace{85pt} (3.9a)\\
where : $2\,{K_{\alpha {\kern 1pt} \beta {\kern 1pt} \varphi {\kern 1pt} \gamma }} = {\widetilde D_\varphi }\overline S {_{\alpha {\kern 1pt} \beta {\kern 1pt} \gamma }} - {\widetilde D_\gamma }\overline S {_{\alpha {\kern 1pt} \beta {\kern 1pt} \varphi }} + \overline S {_{\alpha {\kern 1pt} \delta {\kern 1pt} \varphi }}\,\overline S _{\,.\;\beta {\kern 1pt} \gamma }^\delta  - \overline S {_{\alpha {\kern 1pt} \delta {\kern 1pt} \gamma }}\,\overline S _{\,.\;\beta {\kern 1pt} \varphi }^\delta $
\hspace{2pt} (3.9b)\\
which satisfies the same symmetry relations as the curvature tensor :\\
\hspace*{60pt}
${K_{\alpha {\kern 1pt} \beta {\kern 1pt} \varphi {\kern 1pt} \gamma }} =  - {K_{\beta {\kern 1pt} \alpha {\kern 1pt} \varphi {\kern 1pt} \gamma }}$
\hspace{20pt}
${K_{\alpha {\kern 1pt} \beta {\kern 1pt} \varphi {\kern 1pt} \gamma }} =  - {K_{\alpha {\kern 1pt} \beta {\kern 1pt} \gamma {\kern 1pt} \varphi }}$
\hspace{35pt} (3.9c)\\
Going back to equation (3.5), we define :\\
\hspace*{10pt}
$D_R^{a{\kern 1pt} b{\kern 1pt} c} = {\partial _d}{R^{a{\kern 1pt} b{\kern 1pt} c{\kern 1pt} d}} + \Gamma _{\,.\;e{\kern 1pt} d}^a\,{R^{e{\kern 1pt} b{\kern 1pt} c{\kern 1pt} d}} + \Gamma _{\,.\;e{\kern 1pt} d}^b\,{R^{a{\kern 1pt} e{\kern 1pt} c{\kern 1pt} d}} + \widetilde \Gamma _{\,.\;e{\kern 1pt} d}^c\,{R^{a{\kern 1pt} b{\kern 1pt} e{\kern 1pt} d}} + \widetilde \Gamma _{\,.\;e{\kern 1pt} d}^d\,{R^{a{\kern 1pt} b{\kern 1pt} c{\kern 1pt} e}}$\\
which is also :\\
$D_R^{a{\kern 1pt} b{\kern 1pt} c} = h_\alpha ^a\,h_\beta ^b\,h_\gamma ^c\,$\\
\hspace*{30pt}
$\left[ {{\partial _\delta }{R^{\alpha {\kern 1pt} \beta {\kern 1pt} \gamma {\kern 1pt} \delta }} + \Gamma _{\,.\;\varepsilon {\kern 1pt} \delta }^\alpha \,{R^{\varepsilon {\kern 1pt} \beta {\kern 1pt} \gamma {\kern 1pt} \delta }} + \Gamma _{\,.\;\varepsilon {\kern 1pt} \delta }^\beta \,{R^{\alpha {\kern 1pt} \varepsilon {\kern 1pt} \gamma {\kern 1pt} \delta }}
+ \widetilde \Gamma _{\,.\;\varepsilon {\kern 1pt} \delta }^\gamma \,{R^{\alpha {\kern 1pt} \beta {\kern 1pt} \varepsilon {\kern 1pt} \delta }} + \widetilde \Gamma _{\,.\;\varepsilon {\kern 1pt} \delta }^\delta \,{R^{\alpha {\kern 1pt} \beta {\kern 1pt} \gamma {\kern 1pt} \varepsilon }}} \right]$\\
and finally : \\
$D_R^{\alpha {\kern 1pt} \beta {\kern 1pt} \gamma } = {\widetilde D_\delta }{\widetilde R^{\alpha {\kern 1pt} \beta {\kern 1pt} \gamma {\kern 1pt} \delta }} + {\widetilde D_\delta }{\kern 1pt} {K^{\alpha {\kern 1pt} \beta {\kern 1pt} \gamma {\kern 1pt} \delta }} + \overline S _{\,.\;\varepsilon {\kern 1pt} \delta }^\alpha \,{\widetilde R^{\varepsilon {\kern 1pt} \beta {\kern 1pt} \gamma {\kern 1pt} \delta }} + \overline S _{\,.\;\varepsilon {\kern 1pt} \delta }^\beta \,{\widetilde R^{\alpha {\kern 1pt} \varepsilon {\kern 1pt} \gamma {\kern 1pt} \delta }} + \overline S _{\,.\;\varepsilon {\kern 1pt} \delta }^\alpha \,{K^{\varepsilon {\kern 1pt} \beta {\kern 1pt} \gamma {\kern 1pt} \delta }} $\\
\hspace*{110pt}
$+ \overline S _{\,.\;\varepsilon {\kern 1pt} \delta }^\beta \,{K^{\alpha {\kern 1pt} \varepsilon {\kern 1pt} \gamma {\kern 1pt} \delta }}$
\hspace{110pt} (3.10)\\
This tensor satisfies the anti-symmetry relation :  $D_R^{\alpha {\kern 1pt} \beta {\kern 1pt} \gamma } =  - D_R^{\beta {\kern 1pt} \alpha {\kern 1pt} \gamma }$ as expected.\\

 The equations of motion can now be written with tensor components. The variation of the Lagrangian with respect to the connexion is :\\
$ - \,\frac{1}{{N{\kern 1pt} {\beta ^2}}}\,\frac{1}{{dV}}\,\frac{{\delta L}}{{\delta {\omega _{a{\kern 1pt} b{\kern 1pt} \gamma }}}} = 4{\kern 1pt} \eta \,D_R^{\alpha {\kern 1pt} \beta {\kern 1pt} \gamma }\,h_\alpha ^a\,h_\beta ^b\; + \;h_c^\gamma \,\left( {{\eta ^{b{\kern 1pt} c}}\,\overline S {^a} - {\eta ^{a{\kern 1pt} c}}\,\overline S {^b}} \right)$\\
\hspace*{80pt}
$ + \quad h_c^\gamma \,\left( {2{\kern 1pt} \mu \,\overline S {^{a{\kern 1pt} b{\kern 1pt} c}} + (\mu  - 1)\,(\overline S {^{c{\kern 1pt} a{\kern 1pt} b}} - \,\overline S {^{c{\kern 1pt} b{\kern 1pt} a}})} \right)$
\hspace{25pt} (3.11a)\\
where : $\overline S {^a} = \overline S _{\,.\;{\kern 1pt} .{\kern 1pt} \;d}^{d{\kern 1pt} a}$  . Or, equivalently :\\
$4{\kern 1pt} \eta \,D_R^{\alpha {\kern 1pt} \beta {\kern 1pt} \gamma }\,\; + \;\,\left( {{g^{\beta {\kern 1pt} \gamma }}\,\overline S {^\alpha } - {g^{\alpha {\kern 1pt} \gamma }}\,\overline S {^\beta }} \right) + 2{\kern 1pt} \mu \,\overline S {^{\alpha {\kern 1pt} \beta {\kern 1pt} \gamma }}$\\
\hspace*{100pt}
$ + (\mu  - 1)\,(\overline S {^{\gamma {\kern 1pt} \alpha {\kern 1pt} \beta }} - \,\overline S {^{\gamma {\kern 1pt} \beta {\kern 1pt} \alpha }}) \sim {S_\psi }^{\alpha {\kern 1pt} \beta {\kern 1pt} \gamma }$
\hspace{35pt} (3.11b)\\
where : ${S_\psi }^{\alpha {\kern 1pt} \beta {\kern 1pt} \gamma }$  is the spin tensor of the matter fields.\\

  The variation of the Lagrangian with respect to the fields $\left\{ {h_a^\alpha } \right\}$ is :\\
$\frac{1}{{2\,N{\kern 1pt} {\beta ^2}}}\,\frac{1}{{dV}}\,{\delta _h}L = h_a^\alpha \,\delta h_\gamma ^a\;\{ \;\widetilde G_{\,.\;\alpha }^\gamma  + K_{\,.\;\alpha }^\gamma  - \frac{K}{2}\,\delta _\alpha ^\gamma  - \lambda \,\delta _\alpha ^\gamma \; - \;2{\kern 1pt} \mu \,({\widetilde D_\delta }{\kern 1pt} \overline S _{\alpha \,.\;.}^{\,.\;\delta {\kern 1pt} \gamma } - {\widetilde D_\delta }{\kern 1pt} \overline S _{\alpha \,.\;.}^{\,.\;\gamma {\kern 1pt} \delta })$\\
\hspace*{80pt}
$ - \quad 2{\kern 1pt} \mu \;(\;\overline S {_{\alpha {\kern 1pt} \varepsilon {\kern 1pt} \delta }}\;\overline S {^{\varepsilon {\kern 1pt} \delta {\kern 1pt} \gamma }} - \overline S {_{\alpha {\kern 1pt} \varepsilon {\kern 1pt} \delta }}\;\overline S {^{\varepsilon {\kern 1pt} \gamma {\kern 1pt} \delta }}\,)\;\;\} $
\hspace{47pt} (3.12)\\
where :  $\widetilde G_{\,.\;\alpha }^\gamma  = \widetilde R_{\,.\;\alpha }^\gamma  - \frac{{\widetilde R}}{2}\;\delta _\alpha ^\gamma $ is the Einstein tensor obtained with the Christoffel symbol only, and : ${K_{\gamma {\kern 1pt} \delta }} = K_{\,.\;\gamma {\kern 1pt} \varepsilon {\kern 1pt} {\kern 1pt} \delta }^\varepsilon $ .\\
 The equations (3.11) and (3.12) are nonlinear, as are gauge theory equations. In the next two sections we shall look at these equations for three particular cases : the central symmetric gravitational field, the expanding universe at present time and the asymptotic gravitational field induced by a rotating body, in order to show that they do not contradict the present observations. If the quadratic part of the Lagrangian (3.1) does not exist ( $\eta  = 0$ ), then the equations (3.11) are satisfied if the torsion is null, therefore the equations (3.12) are simply the usual Einstein equations with a cosmological constant. In the case of the Einstein-Hilbert Lagrangian, the torsion is coupled to the spin density directly via a “contact term” and does not propagate. Here, equation (3.11b) shows that torsion can propagate.\\


\section{The static central symmetric case.}

  The solution of the Einstein equations in the static spherical symmetric case is given by the Schwarzchild linear element. Besides torsion terms, the Lagrangian (2.10) contains also a term corresponding to a cosmological constant. We shall proceed as follows : we first consider the linear element which is solution of the Einstein equations with cosmological constant. Then we compute the term : ${\widetilde D_R}^{\alpha {\kern 1pt} \beta {\kern 1pt} \gamma } = {\widetilde D_\delta }{\widetilde R^{\alpha {\kern 1pt} \beta {\kern 1pt} \gamma {\kern 1pt} \delta }}$ of (3.11), and look at the consequences for the torsion.\\

  We start from the following linear element (see for instance \cite{LL}) :\\
\hspace*{50pt}
$d{s^2} = {e^\nu }\,d{t^2} - {e^\sigma }\,d{r^2}\, - {r^2}\,(\,d{\theta ^2} + {\sin ^2}(\theta )\,d{\varphi ^2})$
\hspace{45pt} (4.1)\\
where : $\nu $ and $\sigma $ are functions of the radius : $r$ . We set :\\
\hspace*{30pt}
$c = \cos (\theta )$ \hspace{10pt} $s = \sin (\theta )$ \hspace{10pt} ${c_\varphi } = \cos (\varphi )$ \hspace{10pt} ${s_\varphi } = \sin (\varphi )$\\
  The Einstein equations with cosmological constant are satisfied if :\\
\hspace*{72pt}
$\nu  =  - \sigma  = \ln (1 + \frac{\alpha }{r} - \frac{\lambda }{3}\,{r^2})$
\hspace{105pt} (4.2)\\
(for a discussion of this expression, see for instance \cite{Mathpages}).\\
  The non-zero components of :  ${\widetilde D_R}^{\alpha {\kern 1pt} \beta {\kern 1pt} \gamma }$ (up to the anti-symmetry in the two first indices) are :\\
\hspace*{40pt}
${\widetilde D_R}^{0{\kern 1pt} 1{\kern 1pt} 0} = {e^{ - \nu  - 2{\kern 1pt} \sigma }}\;\left\{ {{\partial _r}\widetilde R_{\,.\;1{\kern 1pt} 0{\kern 1pt} 1}^0 + (\,\frac{2}{r} - \sigma '\,)\,\widetilde R_{\,.\;1{\kern 1pt} 0{\kern 1pt} 1}^0 + \frac{{\nu '}}{{{r^2}}}} \right\}$
\hspace{25pt} (4.3)\\
where the sign prime means the derivation with respect to $r$ , and where :\\
\hspace*{70pt} 
$\widetilde R_{\,.\;1{\kern 1pt} 0{\kern 1pt} 1}^0 =  - \;\frac{{\nu ''}}{2} + \frac{{\nu '}}{4}\;(\,\sigma ' - \nu '\,)$
\hspace{95pt} (4.4)\\
\hspace*{40pt}
${\widetilde D_R}^{1{\kern 1pt} 2{\kern 1pt} 2} = \frac{{{e^{ - 2{\kern 1pt} \sigma }}}}{{2\,{r^3}}}\;\left\{ {\sigma '' + \frac{{\nu '{{\kern 1pt} ^2}}}{{{r^2}}} + \frac{{\nu '\,\sigma '}}{2} - \sigma '{{\kern 1pt} ^2} + \frac{2}{{{r^2}}}} \right\} - \frac{{{e^{ - \sigma }}}}{{{r^5}}}$
\hspace{33pt} (4.5)\\
\hspace*{120pt}
${\widetilde D_R}^{1{\kern 1pt} 3{\kern 1pt} 3} = \frac{1}{{{s^2}}}\;{\widetilde D_R}^{1{\kern 1pt} 2{\kern 1pt} 2}$\\
 These non-zero components are consistent with the constraints (A.7). Now, using the expression (4.2), one gets :
\hspace{10pt}
${\widetilde D_R}^{\alpha {\kern 1pt} \beta {\kern 1pt} \gamma } = {\widetilde D_\delta }{\widetilde R^{\alpha {\kern 1pt} \beta {\kern 1pt} \gamma {\kern 1pt} \delta }} = 0$
\hspace{20pt} (4.6)\\
  The equations (3.11) are satisfied if the contorsion tensor (the torsion tensor) is null. This means that the equations (3.12) reduce to the ordinary Einstein equations, and that (4.1) with (4.2) is a solution of the problem, as in the ordinary case. However, we do not claim that it is the unique solution to equations (3.11) and (3.12).


\section{The expanding isotropic universe.}

  In this section we consider that the global structure of the universe is described by an isotropic space and that the space-time linear element is \cite{LL} :\\
\hspace*{80pt}
$d{\kern 1pt} {s^2} = {a^2}(\eta )\;(\,d{\kern 1pt} {\eta ^2} + \;{\gamma _{\mu {\kern 1pt} \nu }}\;d{x^\mu }d{x^\nu })$
\hspace{63pt} (5.1)\\
where : $\eta  = {x^0}$  is the “conformal” time. We use the standard notation for it, although this symbol has already been used as the coefficient of the quadratic term in the Lagrangian, but there should be no confusion. The metric coefficients : ${\gamma _{\mu {\kern 1pt} \nu }}$ are assumed to be functions of the space coordinates only, and are given, up to a sign,  by (A2) and (A5) for the hyperbolic and spherical cases respectively.
  As usually done, we set : $H=\frac{\dot{a}}{a}$ , where the dot above the letter means the derivative with respect to $\eta$ .\\
The first step is to compute the term : $D_R^{\alpha {\kern 1pt} \beta {\kern 1pt} \gamma }$ of (3.11) using (3.10), as in the previous section. The only non-zero components of $D_R^{\alpha {\kern 1pt} \beta {\kern 1pt} \gamma }$ are :\\
\hspace*{70pt}
$\widetilde D_R^{0{\kern 1pt} \mu {\kern 1pt} \nu } = \frac{1}{{{a^6}}}\;\left[ {\ddot{H} - 2{\kern 1pt} \,H\,(1 + {H^2})} \right]\;{\gamma ^{\mu {\kern 1pt} \nu }}$
\hspace{62pt} (5.2)\\
in agreement with the general form (A9).\\
  Then we have to compute :  $K_{\;.\;\;.\;\;\gamma {\kern 1pt} \delta }^{\alpha {\kern 1pt} \beta }$ . According to (A9) the only non-zero components of the contorsion tensor are :
\hspace*{10pt}
$\overline S {^{0{\kern 1pt} \mu {\kern 1pt} \nu }} = q(\eta )\;{\gamma ^{\mu {\kern 1pt} \nu }}$
\hspace{25pt} (5.3)\\
up to the anti-symmetry on the first two indices. In (5.1), $\eta$ and ${x^\mu }$ have no dimension, then $\widetilde \Gamma _{\,.\;\beta {\kern 1pt} \gamma }^\alpha $  has no dimension, therefore $q$ must have the dimension of a length to the power of -4. Using (5.3), the non-zero components of : $K_{\;.\;\;.\;\;\gamma {\kern 1pt} \delta }^{\alpha {\kern 1pt} \beta }$  are (no sum over $\mu$ ) :\\  
\hspace*{70pt}
$K_{\;.\;\,.\,\;0\,\mu }^{0{\kern 1pt} \mu } = {a^2}\,( \dot{q}+ 4{\kern 1pt} H{\kern 1pt} q)\quad \forall \,\mu $\\
\hspace*{40pt}
$K_{\;.\;\,.\,\;\rho \,\eta }^{\mu {\kern 1pt} \nu } = {a^2}q\;(\,2\,H - {a^4}q\,)\;(\,\delta _{{\kern 1pt} \rho }^\mu \,\delta _{{\kern 1pt} \eta }^\nu \, - \delta _{{\kern 1pt} \eta }^\mu \,\delta _{{\kern 1pt} \rho }^\nu \,)$
\hspace{63pt} (5.4)\\
and finally, the only components of ${\widetilde D_\delta }{\kern 1pt} {K^{\alpha {\kern 1pt} \beta {\kern 1pt} \gamma {\kern 1pt} \delta }}={\widetilde D}_{K} ^{\alpha {\kern 1pt} \beta {\kern 1pt} \gamma}$  left are :\\
\hspace*{10pt}
${\widetilde D_K}^{0{\kern 1pt} \mu {\kern 1pt} \nu }  =  - \;\frac{1}{{{a^4}}}\;\left[ {\partial _\eta ^2({a^2}q) + 4H\,{\partial _\eta }({a^2}q) + 2\,\dot{H}{a^2}q + 2{\kern 1pt} H\,{a^6}{q^2}} \right]\;{\gamma ^{\mu {\kern 1pt} \nu }}$
\hspace{7pt} (5.5)\\
Gathering all the terms of  (3.11b), one obtains :\\
\hspace*{40pt}
$ D_R^{0{\kern 1pt} \mu {\kern 1pt} \nu } = \widetilde D_R^{0{\kern 1pt} \mu {\kern 1pt} \nu } + {\widetilde D_K}^{0{\kern 1pt} \mu {\kern 1pt} \nu } + q\,{a^4}{\kern 1pt} {\gamma _{\rho {\kern 1pt} \eta }}\,{\widetilde R^{\rho {\kern 1pt} \mu {\kern 1pt} \nu {\kern 1pt} \eta }} + q\,{a^4}{\kern 1pt} {\gamma _{\rho {\kern 1pt} \eta }}\,{K^{\rho {\kern 1pt} \mu {\kern 1pt} \nu {\kern 1pt} \eta }}$\\
with : \hspace{30pt} ${\widetilde R^{\rho {\kern 1pt} \mu {\kern 1pt} \nu {\kern 1pt} \eta }} = \frac{1}{{{a^6}}}\;\left( {{H^2} - \varepsilon } \right)\;\left( {{\gamma ^{\mu {\kern 1pt} \nu }}{\kern 1pt} {\gamma ^{\rho {\kern 1pt} \eta }} - {\gamma ^{\rho {\kern 1pt} \nu }}{\kern 1pt} {\gamma ^{\mu {\kern 1pt} \eta }}} \right)$ \\
and where : $\epsilon=1$ for a spherical space, $\epsilon=-1$ for the hyperbolic case  and  $\epsilon=0$ for  an Euclidean space.\\

 With all the above expressions, and assuming that there is no macroscopic spin density, equation (3.11b) becomes :\\
\hspace*{20pt}
$4{\kern 1pt} \eta \,\left\{ {{{\widetilde D}_R}^{0{\kern 1pt} \mu {\kern 1pt} \nu }\, + {{\widetilde D}_K}^{0{\kern 1pt} \mu {\kern 1pt} \nu } + 2\,\frac{q}{{{a^2}}}\,({H^2} - \varepsilon )\,{\gamma ^{\mu {\kern 1pt} \nu }} - 2\,{a^2}{q^2}{\kern 1pt} (2H - {a^4}q{\kern 1pt} )\,{\gamma ^{\mu {\kern 1pt} \nu }}} \right\}$\\
\hspace*{120pt}
$ + \;(\mu  - 2)\,q\,{\gamma ^{\mu {\kern 1pt} \nu }} = 0$
\hspace{90pt} (5.6)\\

  In order to discuss the Einstein equations we need the following results :\\
\hspace*{20pt}
${a^2}\,\widetilde G_{\,.\;0}^0 = 3{H^2} + 3\,\varepsilon $
\hspace{20pt}
$\widetilde G_{\,.\;\mu }^0 = 0$
\hspace{20pt}
${a^2}\,\widetilde G_{\,.\;\nu }^\mu  = (\varepsilon  + {H^2} + 2\,\dot{H})\,\delta _{{\kern 1pt} \nu }^\mu $\\
\hspace*{80pt}
$K_{\,.\;0}^0 - \frac{K}{2}\, = 3\,{a^2}q\;({a^4}q - 2H)$\\
\hspace*{60pt}
$K_{\,.\;\mu }^\mu  - \frac{K}{2}\, = \,{a^2}q\;(\,{a^4}q - 10H - 2\,\frac{\dot{q}}{q}\,)\quad \forall \;\mu $ \\
In equation (3.12) one has to compute : ${E^{\alpha {\kern 1pt} \gamma }} = {\widetilde D_\delta }{\kern 1pt} {\Omega ^{\alpha {\kern 1pt} \delta {\kern 1pt} \gamma }} - {\widetilde D_\delta }{\kern 1pt} {\Omega ^{\alpha {\kern 1pt} \gamma {\kern 1pt} \delta {\kern 1pt} }}$ for the different cases. We obtain :\\
\hspace*{30pt}
${E^{0{\kern 1pt} 0}} =  - 3\,H{\kern 1pt} q$
\hspace{10pt}
${E^{0{\kern 1pt} \mu }} = 0$
\hspace{10pt}
${E^{\mu {\kern 1pt} \nu }} =  - (\dot{q} + 5\,H{\kern 1pt} q)\;{\gamma ^{\mu {\kern 1pt} \nu {\kern 1pt} }}$
\hspace{30pt} \\
Finally, calling $T_{\;\gamma }^\alpha $ the energy-momentum tensor, the Einstein equations (3.12) are :
\hspace{30pt}
$\widetilde R_{\,.\;0}^0 - \frac{{\widetilde R}}{2}\; - \lambda  - 12\,H{\kern 1pt} {a^2}{\kern 1pt} q + 9\,{a^6}{\kern 1pt} {q^2} = T_{\,.\;0}^0$ \\
\hspace*{25pt}
$\left[ {\widetilde R_{\,.\;\nu }^\mu  - \frac{{\widetilde R}}{2}\,\delta _{{\kern 1pt} \nu }^\mu \; - \lambda \,\delta _{{\kern 1pt} \nu }^\mu } \right] - \left[ {20\,H{\kern 1pt} {a^2}{\kern 1pt} q + 4\,{a^2}{\kern 1pt} \dot{q} - \,{a^6}{\kern 1pt} {q^2}} \right]\,\delta _{{\kern 1pt} \nu }^\mu  = T_{{\kern 1pt} \nu }^\mu $
\hspace{10pt} (5.7)\\
and equation (3.12) is null if the indices are : $\alpha  = 0\;,\;\gamma  \equiv \mu $ or the converse.\\

 Are equations (5.6) and (5.7) compatible with the present day observable universe ? 
 The only length in the problem is $a(\eta)$ in (5.1), therefore we set : $q = r\,{a^{ - {\kern 1pt} 4}}$ , which gives :
\hspace{10pt}
${\widetilde D_K}^{0{\kern 1pt} \mu {\kern 1pt} \nu } =  - \;\frac{1}{{{a^6}}}\;\left[ {\ddot{r} - 4{H^2}{\kern 1pt} r + 2{\kern 1pt} H{\kern 1pt} {r^2}} \right]\;{\gamma ^{\mu {\kern 1pt} \nu }}$ \\
Neglecting the terms $\ddot{H}$ and $H^2$ in (5.2), equation (5.6) becomes :\\
$\frac{{4{\kern 1pt} \eta }}{{{a^2}}}\,\left\{ {2H + \ddot{r} - 4{H^2}{\kern 1pt} r + 2{\kern 1pt} H{\kern 1pt} {r^2}\, - 2{\kern 1pt} r\,({H^2} - \varepsilon ) + 2\,{r^2}(2H - r)} \right\} + \;(\mu  - 2)\,r = 0$
\hspace{10pt}\\
When : $a \to \infty $ , this equation implies : $r \to 0$ . Neglecting again the terms $H^2$ and $r^2$ , one has, at lowest order, when $a \to \infty $ :  $r \sim  - \,\frac{{4{\kern 1pt} \eta {\kern 1pt} H}}{{{a^2}}}\,$ \\     
 This asymptotic value brings, in (5.7), corrections of order $a^{-2}$ with respect to the usual Einstein equations without torsion. Therefore, when the universe has become large, the torsion is invisible.


\section{The asymptotic gravitational field of a rotating body.}

 The space-time linear element induced by a rotating body, far from it, is written as \cite{ABS} :\\
\hspace*{20pt}
$d{s^2} = a\,d{t^2} - b\,d{r^2}\, - {r^2}\,(\,d{\theta ^2} + {\sin ^2}(\theta )\,d{\varphi ^2}) - \,\frac{{2\,l}}{r}\,{s^2}\,dt\,d\varphi $
\hspace{25pt} (6.1)\\
where $l$ is proportional to the source body angular momentum, and where :\\
\hspace*{30pt}
${g_{0\,0}} = a = 1 + \frac{\alpha }{r} + O({r^{ - 2}})$
\hspace{20pt}
$ - {g_{1\,1}} = b = \frac{1}{a} + O({r^{ - 2}})$
\hspace{30pt} \\
with $\alpha$ a constant. In this section, the cosmological constant is set to 0 .
The non zero Christoffel symbols at leading order and at lowest order in $l$ are :\\
$\widetilde \Gamma {\kern 1pt} _{\,.\;0{\kern 1pt} 1}^0 \approx \frac{{a'}}{{2\,a}}$
\hspace{12pt}
$\widetilde \Gamma {\kern 1pt} _{\,.\;1{\kern 1pt} 3}^0 \approx \frac{{3\,l\,{s^2}}}{{2\,a\,{r^2}}}$
\hspace{12pt}
$\widetilde \Gamma {\kern 1pt} _{\,.\;1{\kern 1pt} 1}^1 = \frac{{b'}}{{2\,b}}$
\hspace{12pt}
$\widetilde \Gamma {\kern 1pt} _{\,.\;0{\kern 1pt} 0}^1 = \frac{{a'}}{{2\,b}}$
\hspace{12pt}
$\widetilde \Gamma {\kern 1pt} _{\,.\;0{\kern 1pt} 3}^1 = \frac{{l\,{s^2}}}{{2\,b\,{r^2}}}$
\hspace{10pt} \\
\hspace*{10pt}
$\widetilde \Gamma {\kern 1pt} _{\,.\;2{\kern 1pt} 2}^1 =  - \frac{r}{b}$
\hspace{10pt}
$\widetilde \Gamma {\kern 1pt} _{\,.\;3{\kern 1pt} 3}^1 =  - \frac{{r\,{s^2}}}{b}$
\hspace{10pt}
$\widetilde \Gamma {\kern 1pt} _{\,.\;0{\kern 1pt} 3}^2 =  - \frac{{l\,s{\kern 1pt} c}}{{{r^3}}}$
\hspace{15pt}
$\widetilde \Gamma {\kern 1pt} _{\,.\;1{\kern 1pt} 2}^2 = \frac{1}{r}$
\hspace{17pt} (6.2)\\
\hspace*{50pt}
$\widetilde \Gamma {\kern 1pt} _{\,.\;3{\kern 1pt} 3}^2 =  - s{\kern 1pt} c$
\hspace{12pt}
$\widetilde \Gamma {\kern 1pt} _{\,.\;0{\kern 1pt} 1}^3 \approx  - \frac{{l\,}}{{2{\kern 1pt} {r^3}}}\;\left( {\frac{{a'}}{a} + \frac{1}{r}} \right)$\\
\hspace*{50pt}
$\widetilde \Gamma {\kern 1pt} _{\,.\;0{\kern 1pt} 2}^3 \approx \frac{{l{\kern 1pt} c}}{{{r^3}s}}$
\hspace{12pt}
$\widetilde \Gamma {\kern 1pt} _{\,.\;1{\kern 1pt} 3}^3 \approx \frac{1}{r}$
\hspace{12pt}
$\widetilde \Gamma {\kern 1pt} _{\,.\;2{\kern 1pt} 3}^3 = \frac{c}{s}$ \\
The components of the Ricci tensor built with the Christoffel symbols are :\\
\hspace*{30pt}
$\widetilde R{\kern 1pt} _{\;0}^0 = \frac{1}{{2\,b}}\;\left[ {\frac{{a''}}{a} - \frac{{a'{\,^2}}}{{2\,{a^2}}} + \frac{{a'}}{a}\,(\,\frac{a}{r} - \frac{{b'}}{{2\,b}}\,)} \right] + O({r^{ - 6}})$ \\
\hspace*{30pt}
$\widetilde R{\kern 1pt} _{\;1}^1 = \frac{1}{{2\,b}}\;\left[ { - \frac{{a''}}{a} + \frac{{a'{\,^2}}}{{2\,{a^2}}} + \frac{{a'\,b'}}{{2\,a\,b}}\,} \right] + \frac{{b'}}{{b{\kern 1pt} r}} + O({r^{ - 6}})$ \\
\hspace*{30pt}
$\widetilde R{\kern 1pt} _{\;2}^2 = \widetilde R{\kern 1pt} _{\;3}^3 = \,\frac{1}{{2\,b{\kern 1pt} r}}\;\left[ {\frac{{a'}}{a} - \frac{{b'}}{b}\,} \right] - \frac{1}{{{r^2}}}\,(1 - \frac{1}{b}) + O(\frac{{{l^2}}}{{{r^6}}})$\\
\hspace*{30pt}
${R_{0{\kern 1pt} 1}} = {R_{0{\kern 1pt} 2}} = {R_{1{\kern 1pt} 3}} = {R_{2{\kern 1pt} 3}} = 0$
\hspace{30pt}
${R_{1{\kern 1pt} 2}} = O(\frac{l}{{{r^5}}})$\\
\hspace*{30pt}
$\widetilde R{{\kern 1pt} _{0{\kern 1pt} 3}}{\kern 1pt}  = \widetilde R{{\kern 1pt} _{3{\kern 1pt} {\kern 1pt} 0}}{\kern 1pt}  = \,\frac{{l\,{s^2}}}{{{r^3}}}\;\left[ {1 - \frac{1}{b} - \frac{{5{\kern 1pt} a'{\kern 1pt} r}}{{4{\kern 1pt} a{\kern 1pt} b}} - \frac{{b'{\kern 1pt} r}}{{4{\kern 1pt} {b^2}}}\,} \right] = O(\frac{l}{{{r^4}}})$\\
The important point is that the diagonal terms of the Ricci tensor components due to the Christoffel symbols are unchanged with respect to the spherical symmetric case at lowest order, and that these diagonal terms are of order : $r^{-3}$ .\\
   As before the next step is to compute the quantities :  ${\widetilde D_R}^{\alpha {\kern 1pt} \beta {\kern 1pt} \gamma } = {\widetilde D_\delta }{\widetilde R^{\alpha {\kern 1pt} \beta {\kern 1pt} \gamma {\kern 1pt} \delta }}$   which appear in equations (3.10) and (3.11). The calculations are done at lowest order only, which means that we write :  $\widetilde \Gamma _{\,.\;\beta {\kern 1pt} \gamma }^\alpha  = \mathop {\widetilde \Gamma }\limits_0 {\kern 1pt} _{\,.\;\beta {\kern 1pt} \gamma }^\alpha  + \mathop {\widetilde \Gamma }\limits_1 {\kern 1pt} _{\,.\;\beta {\kern 1pt} \gamma }^\alpha $ , where the first term on the right corresponds to the case : $l=0$ , and where the second term are of order : $l$ . Likewise, we write : $\widetilde R{{\kern 1pt} ^{\alpha {\kern 1pt} \beta {\kern 1pt} \gamma {\kern 1pt} \delta }} = \mathop {\widetilde R}\limits_0 {{\kern 1pt} ^{\alpha {\kern 1pt} \beta {\kern 1pt} \gamma {\kern 1pt} \delta }} + \mathop {\widetilde R}\limits_1 {{\kern 1pt} ^{\alpha {\kern 1pt} \beta {\kern 1pt} \gamma {\kern 1pt} \delta }}$  with the same meaning. In the calculations of the covariant derivatives, terms like : $\mathop {\widetilde \Gamma }\limits_1 {\kern 1pt} _{\,.\;\varepsilon {\kern 1pt} \delta }^\alpha \,\mathop {\widetilde R}\limits_1 {{\kern 1pt} ^{\varepsilon {\kern 1pt} \beta {\kern 1pt} \gamma {\kern 1pt} \delta }}$ are neglected. The result is that the non-zero terms are :\\
${\widetilde D_R}^{0{\kern 1pt} 1{\kern 1pt} 3} \approx \frac{{3\,l}}{{b{\kern 1pt} {r^6}}}\,(\,\frac{{{c^2}}}{{{s^2}}} - \frac{1}{2}\,) + O(\frac{1}{{{r^7}}})$
\hspace{10pt}
${\widetilde D_R}^{0{\kern 1pt} 3{\kern 1pt} 1} \approx O(\frac{l}{{{r^6}}})$
\hspace{10pt}
${\widetilde D_R}^{1{\kern 1pt} 3{\kern 1pt} 0} \approx \frac{{3\,l}}{{b{\kern 1pt} {r^6}}}\,(2 - \,\frac{{{c^2}}}{{2\,{s^2}}}\,) + O(\frac{1}{{{r^7}}})$\\
\hspace*{10pt}
${\widetilde D_R}^{0{\kern 1pt} 2{\kern 1pt} 3} \approx \frac{{l\,c}}{{{r^6}s}}\,(\,\frac{9}{r} - \frac{{a'}}{a}\,)$
\hspace{10pt}
${\widetilde D_R}^{0{\kern 1pt} 3{\kern 1pt} 2} \approx  - {\widetilde D_R}^{0{\kern 1pt} 2{\kern 1pt} 3}$
\hspace{10pt}
${\widetilde D_R}^{2{\kern 1pt} 3{\kern 1pt} 0} \approx  - 2\,{\widetilde D_R}^{0{\kern 1pt} 2{\kern 1pt} 3}$
\hspace{22pt} (6.3)\\
 This is consistent with the fact that : ${\widetilde D_R}^{\alpha {\kern 1pt} \beta {\kern 1pt} \gamma } = 0$ when : $l=0$ (section 4) .\\
From equation (3.11) this means that the contorsion has non-zero components. In each of these non-zero terms the indices are all different. In other words, the quantities : ${\widetilde D_R}^{\alpha {\kern 1pt} \beta {\kern 1pt} \gamma }$ are null if two of the indices are equal. If we look back at the Christoffel symbols, the non-zero terms in (6.3) bear the same triplet of indices as the Christoffel symbols of order $l$ . This suggests to try a solution of the form :
\hspace{40pt}
$\overline S {_{\alpha {\kern 1pt} \beta {\kern 1pt} \gamma }} = \frac{{{\omega _{\alpha {\kern 1pt} \beta {\kern 1pt} \gamma }}(r,\theta )}}{{{r^p}}}$
\hspace{10pt}  if : \hspace{10pt}
$\left\{ {\alpha ,\,\beta ,\,\gamma } \right\} = \left\{ {0,\,1,\,3} \right\}$
\hspace{35pt} (6.4)\\
\hspace*{50pt}
$\overline S {_{\alpha {\kern 1pt} \beta {\kern 1pt} \gamma }} = \frac{{{\omega _{\alpha {\kern 1pt} \beta {\kern 1pt} \gamma }}(r,\theta )}}{{{r^q}}}$
\hspace{10pt}  if : \hspace{10pt}
$\left\{ {\alpha ,\,\beta ,\,\gamma } \right\} = \left\{ {0,\,2,\,3} \right\}$ \\
In the following calculations, the indices are raised or lowered using only the diagonal terms of the metric tensor, because the non-diagonal terms introduce higher order contributions.\\
  The hypotheses (6.4) simplify the evolution equations. Firstly, since all the indices are different in (6.4), one has :  $\overline S {^\alpha } = 0$  . Secondly, when two of the indices of the triplet $\left\{ {\alpha ,\,\beta ,\,\gamma } \right\}$ are equal, equation (3.11b) (in vacuum) reduces to the condition : ${D_R}^{\alpha {\kern 1pt} \beta {\kern 1pt} \gamma } = 0$  .\\
 Now we consider the fifth term of Einstein equation (3.12) :\\
\hspace*{10pt}
${\widetilde D_\delta }{\kern 1pt} \overline S _{\alpha \,.\;.}^{\,.\;\delta {\kern 1pt} \gamma } - {\widetilde D_\delta }{\kern 1pt} \overline S _{\alpha \,.\;.}^{\,.\;\gamma {\kern 1pt} \delta } = {\partial _\delta }{\kern 1pt} \overline S _{\alpha \,.\;.}^{\,.\;\delta {\kern 1pt} \gamma } - {\partial _\delta }{\kern 1pt} \overline S _{\alpha \,.\;.}^{\,.\;\gamma {\kern 1pt} \delta } - \widetilde \Gamma _{\,.\;\alpha {\kern 1pt} \delta }^\varepsilon \,(\overline S _{\varepsilon \,.\;.}^{\,.\;\delta {\kern 1pt} \gamma } - \overline S _{\varepsilon \,.\;.}^{\,.\;\gamma {\kern 1pt} \delta })$\\
\hspace*{85pt}
$ + \;\;\widetilde \Gamma _{\,.\;\varepsilon {\kern 1pt} \delta }^\delta \,(\overline S _{\alpha \,.\;.}^{\,.\;\varepsilon {\kern 1pt} \gamma } - \overline S _{\alpha \,.\;.}^{\,.\;\gamma {\kern 1pt} \varepsilon }) + \widetilde \Gamma _{\,.\;\varepsilon {\kern 1pt} \delta }^\gamma \,(\overline S _{\alpha \,.\;.}^{\,.\;\delta {\kern 1pt} \varepsilon } - \overline S _{\alpha \,.\;.}^{\,.\;\varepsilon {\kern 1pt} \delta })$ \\
If : $\alpha=\gamma$ , this term is null, either because : $\overline S _{\alpha \,.\;.}^{\,.\;\beta {\kern 1pt} \gamma } = 0$ , or because it contains the products of the symmetric Christoffel symbol times an anti-symmetric term.\\
 The hypotheses (6.4) provide a solution to the equation of motion. The calculations are long and will not be detailed, only the conclusions are given. First we consider equation (3.11b) when 2 indices of the triplet $\left\{ {\alpha ,\beta ,\gamma } \right\}$ are equals. We have seen that (3.11b) reduces to : ${D_R}^{\alpha {\kern 1pt} \beta {\kern 1pt} \gamma } = 0$ . The contorsion terms of (3.10) can compensate for the second order terms of ${g_{0{\kern 1pt} 0}}$  and ${g_{1{\kern 1pt} 1}}$ if : $p \ge 1$ and : $q \ge 0$ . Then, when the triplet $\left\{ {\alpha ,\beta ,\gamma } \right\}$ is equal to the sets defined in (6.4), the contorsion terms of (3.10) are of the same order of those of the expressions (6.3) if : $p \ge 2$ and : $q \ge 1$ . But, the left side last two terms of (3.11b) are of the same order of the terms (6.3) if : $p \ge 4$ and :  $q \ge 3$ . The six functions of (6.4) provides enough freedom to cancel the six terms (6.3).

 Conclusion :  the equations of motion (3.11) and (3.12)  can be satisfied if :\\
\hspace*{10pt}
$\overline S {_{0{\kern 1pt} 1{\kern 1pt} 3}}\;,\;\overline S {_{0{\kern 1pt} 3{\kern 1pt} 1}}\;,\;\overline S {_{1{\kern 1pt} 3{\kern 1pt} 0}}\; \sim O(\,\frac{l}{{{r^4}}}\,)$
\hspace{10pt}
$\overline S {_{0{\kern 1pt} 2{\kern 1pt} 3}}\;,\;\overline S {_{0{\kern 1pt} 3{\kern 1pt} 2}}\;,\;\overline S {_{2{\kern 1pt} 3{\kern 1pt} 0}}\; \sim O(\,\frac{l}{{{r^3}}}\,)$
\hspace{12pt} (6.5)\\
Going back to the Christoffel symbols (6.2), these contorsion components bring second order corrections to them and would have no effects on the trajectories of test particles. At last, the contorsion terms (at lowest order) are proportional to the angular momentum as expected.\\


\section*{Appendix A . Spaces of constant curvature, isotropy and invariances.}

  Spaces of constant curvature  \cite{Wolf}  are isotropic spaces and symmetric spaces. This appendix recalls the definition of hyperbolic and spherical spaces and their invariance properties. The 3-dimensional spaces ${H^3}$ and ${S^3}$ are respectively hyper-spheres in Minkowski space ${M^4}$ and Euclidean space ${E^4}$ . The invariance with respect to the rotation group in these larger spaces implies local rotational invariance with respect to a point and invariance by transvections  \cite{Wolf}  which are defined below. 

\subsection{Hyperbolic spaces.}

 The hyperbolic n dimensional space ${H^n}$ is defined as the « upper part » of the sphere of radius $\sqrt {\left| K \right|}$ in the Minkowski space ${M^{n + 1}}$ . More precisely, if $\left\{ {{x^\alpha }} \right\}$ are Cartesian coordinates in  ${M^{n + 1}}$ with origin ${O_M}$ , ${H^n}$  is the surface defined by :
\hspace{80pt}
$\sum\limits_{\alpha  = 0}^{n - 1} {\;{x^\alpha }\,{x^\alpha }\; - \;{x^n}\,{x^n}\;\; = \,\;K}$\\
 where : $K < 0$ and : ${x^n} \ge \sqrt {\left| K \right|}$ . We set : $R = \sqrt {\left| K \right|} $ and use  ‘spherical’ coordinates. For ${H^3}$ in ${M^4}$ :
${x^0}\, = \,R\,sh\chi c\quad , \quad c = \cos (\theta )\quad ,\quad s = \sin (\theta )$\\
\hspace*{35pt}      
 ${x^1}\, = \,R\,sh\chi \,s\,{c_\varphi }\quad \quad ;\quad \quad {c_\varphi } = \cos (\varphi )\quad ,\quad {s_\varphi } = \sin (\varphi )$      \hspace{20pt} (A.1) \\
\hspace*{85pt}
 ${x^2}\, = \,R\,sh\chi \;s\,{s_\varphi }$ \hspace{20pt} ${x^3}\, = \,R\,ch\chi $ \\
\hspace*{75pt}
$\chi  \ge 0$   \hspace{20pt}  $\theta  \in \left[ {0\,,\;\pi } \right]$  
\hspace{20pt}  $\varphi  \in \left[ {0\,,\;2\,\pi } \right]$ \\
Then the linear element of ${H^3}$ is :\\
\hspace*{20pt}
$d{s^2} = {(d{x^0})^2} + {(d{x^1})^2} + {(d{x^2})^2} - {(d{x^3})^2} $\\
\hspace*{75pt}
$= {R^2}\,\left[ {d{\chi ^2} + s{h^2}\chi \;(\,{{(d\theta )}^2} + {s^2}{{(d\varphi )}^2}{\kern 1pt} )} \right]$ 
\hspace{55pt} (A.2)\\
The coordinates $(\chi ,\;\theta ,\;\varphi )$ are the Riemann normal (spherical) coordinates with origin at $\chi  = 0$ , which corresponds to the point $(0,\;0,\;0,\;R)$ in ${M^4}$ . The curvature tensor is : ${R_{\alpha {\kern 1pt} \beta {\kern 1pt} \gamma {\kern 1pt} \delta }} =  - \frac{1}{{{R^2}}}\,({g_{\alpha {\kern 1pt} \gamma }}{\kern 1pt} {g_{\beta {\kern 1pt} \delta }} - {g_{\alpha {\kern 1pt} \delta }}{\kern 1pt} {g_{\beta {\kern 1pt} \gamma }})$
  , the  Ricci tensor is  : ${R_{\alpha {\kern 1pt} \beta }} =  - \frac{2}{{{R^2}}}\,{g_{\alpha {\kern 1pt} \beta }}$  , and the scalar curvature : ${R_H} =  - \frac{6}{{{R^2}}}$  . $R$  is a scale factor, in the following it is set to 1.\\

 ${H^n}$ , which is a space of constant curvature, is a symmetric space. A transvection in a symmetric space is an isometry which generalizes the notion of translation in Euclidean space. It is defined as the product of two successive symmetries with respect to two different points $A$ and $B$ . The geodesic going through these two points is invariant and is called the base geodesic. In ${H^3}$ one can perform a rotation around this base geodesic, it commutes with the transvection and the base geodesic is invariant. The base geodesic of a transvection $\gamma$ is given by the intersection of the invariant plane, associated to the real eigenvalues of the $SO(3,1)$ element representing $\gamma$ in ${M^4}$ , with ${H^3}$ .
  If we call $L$ the « length » of the transvection, which is twice the distance between $A$ and $B$ , a point $p$ whose spherical coordinates are : $(\chi ,\,\theta ,\,\varphi )$ is transformed, by a transvection along $Oz$ , into a point $q$ of coordinates : $({\chi _q},\,\;{\theta _q},\;{\varphi _q})$  given by :\\
\hspace*{70pt}
$ch(\chi_q) = ch(\chi )\,ch(L) + sh(\chi )\,sh(L)\,c$\\
\hspace*{70pt}
$c_q = (ch(\chi )\,sh(L) + c\;sh(\chi )\,ch(L))/\,sh(\chi_q)$
\hspace{40pt} (A.3) \\
\hspace*{130pt}
$\varphi_q = \varphi $\\
where : $c_q=\cos(\theta_q)$ .

\subsection{Spherical spaces.}

 The spherical n dimensional space ${S^n}$ is defined as the sphere of radius $R$ in the Euclidean space ${E^{n + 1}}$ . For ${S^3}$ in $E^4$ :
${x^0}\, = \,R\,\sin \chi \,c$\\
\hspace*{30pt}
${x^1}\, = \,R\,\sin \chi \;s\,{c_\varphi }$
\hspace{10pt}
${x^2}\,\; = \;\,R\;\sin \chi \,s\,{s_\varphi }$
\hspace{10pt}
${x^3}\, = \,R\,\cos \chi $
\hspace{13pt} (A.4) \\
\hspace*{80pt}
$\chi \,,\;\theta  \in \left[ {0\,,\;\pi } \right]$
\hspace{30pt}
$\varphi  \in \left[ {0\,,\;2\,\pi } \right]$\\
Then the linear element of $S^3$ is :\\
\hspace*{40pt}
$d{s^2} = {(d{x^0})^2} + {(d{x^1})^2} + {(d{x^2})^2} + {(d{x^3})^2}$\\
\hspace*{70pt}
$ = {R^2}\,\left[ {d{\chi ^2} + {{\sin }^2}\chi \;(\,{{(d\theta )}^2} + {s^2}{{(d\varphi )}^2}{\kern 1pt} )} \right]$
\hspace{57pt} (A.5) \\
As above $R$ is set to 1. With the same notations as in the hyperbolic case, a point of coordinates :
 $(\chi ,\,\;\theta ,\;\varphi )$ 
is transformed , by a transvection along $Oz$ , into a point whose coordinates are given by :\\
\hspace*{60pt}
$\cos (\chi_q) = \cos (\chi )\,\cos (L) - \sin (\chi )\,\sin (L)\,{\mathop{\rm c}\nolimits} $\\
\hspace*{50pt}
$c_q = (\cos(\chi )\,\sin(L) + c\;\sin(\chi )\,\cos(L))/\,\sin(\chi_q)$
\hspace{40pt} (A.6) \\
\hspace*{130pt}
$\varphi_q = \varphi $
\hspace{10pt} \\

\subsection{Rotational  invariance.}

  In order to obtain the transformation law of tensors, we need to know how local frames transform. With the above Riemann normal coordinates the problem is the same as in Euclidean space. We shall use local orthonormal frames whose basis vectors are aligned with the vectors of the natural frame associated with the usual spherical coordinates in the 3 dimensional Euclidean space. These vectors, named $\overrightarrow {{h_1}} (x),\;\overrightarrow {{h_2}} (x),\;\overrightarrow {{h_3}} (x)$ , are respectively aligned with $\overrightarrow {{e_r}} (x),\;\overrightarrow {{e_\theta }} (x),\;\overrightarrow {{e_\varphi }} (x)$ . The Riemann normal coordinates are defined by :\\ 
\hspace*{80pt}
$x = \chi \;s\,{c_\varphi }$
\hspace{10pt}
$y = \chi \;s\,{s_\varphi }$
\hspace{10pt}
$z = \chi \;c$
\hspace{10pt} \\

  Rotation around the Oz axis : $\chi_q =\chi$ , $\theta_q = \theta$ , $\varphi_q = \varphi+\alpha$ \\
 In that case the transformed local frame of point $p$ coincides with the local frame  at $q$ : 
$\overrightarrow {f{_a}} (q) = \overrightarrow {{h_a}} (q)$  .
 Invariance with respect to these rotations does not say much except that tensors do not depend on $\varphi$ .\\
  Rotation around the Ox axis : the action of a rotation of angle $\alpha$ is written using  Cartesian coordinates :\\
$x_q = x \quad  \to \quad s_q \,c_{q \varphi } = s\,{c_\varphi }$
\hspace{10pt}
$y_q = {c_\alpha }{\kern 1pt} y + {s_\alpha }\,z\quad  \to \quad s_q \,s_{q \varphi } = {c_\alpha }{\kern 1pt} s\,{s_\varphi } + {s_\alpha }\,c$ \\
\hspace*{40pt}
$z_q =  - {s_\alpha }{\kern 1pt} y + {c_\alpha }\,z\quad  \to \quad c_q =  - {s_\alpha }{\kern 1pt} s\,{s_\varphi } + {c_\alpha }\,c$ \\
\hspace*{40pt}
${s_q}\,d{\theta _q} = ({s_\alpha }{\kern 1pt} c\,{s_\varphi } + {c_\alpha }\,s)\;d\theta  + {s_\alpha }{\kern 1pt} s\,{c_\varphi }\;d\varphi $\\
\hspace*{40pt}
$s_q^2\,d\varphi_q =  - {s_\alpha }\,{c_\varphi }\;d\theta  + ({c_\alpha }\,{s^2} + {s_\alpha }{\kern 1pt} s\,c\,{s_\varphi })\;d\varphi $ \\
We get : $\overrightarrow {f{_1}} (q) = \overrightarrow {{h_1}} (q)$  , \hspace{5pt}
$\overrightarrow {f{_2}} (q) = a\,\overrightarrow {{h_2}} (q) + b\,\overrightarrow {{h_3}} (q)$  , \hspace{5pt} $\overrightarrow {f{_3}} (q) = a\,\overrightarrow {{h_3}} (q) - b\,\overrightarrow {{h_2}} (q)$ \\
where : $a = ({c_\alpha }{\kern 1pt} s + {s_\alpha }{\kern 1pt} c\,{s_\varphi })/{s_q}$  ,
  $b =  - {s_\alpha }{\kern 1pt} {c_\varphi }/{s_q}$ \\

The contorsion tensor satisfies : ${\overline S ^{\alpha {\kern 1pt} \beta {\kern 1pt} \gamma }} =  - \,{\overline S ^{\beta {\kern 1pt} \alpha {\kern 1pt} \gamma }}$ . With these transformation laws, we deduce that the only nonzero components of the contorsion tensor are :\\
\hspace*{50pt}
${\overline S _h}^{0{\kern 1pt} 1{\kern 1pt} 0} =  - \,{\overline S _h}^{1{\kern 1pt} 0{\kern 1pt} 0}$
\hspace{50pt}
${\overline S _h}^{1{\kern 1pt} 0{\kern 1pt} 1} =  - \,{\overline S _h}^{0{\kern 1pt} 1{\kern 1pt} 1}$
\hspace{30pt} (A.7) \\
\hspace*{80pt}
${\overline S _h}^{2{\kern 1pt} 0{\kern 1pt} 2} =  - \,{\overline S _h}^{0{\kern 1pt} 2{\kern 1pt} 2} = {\overline S _h}^{3{\kern 1pt} 0{\kern 1pt} 3} =  - \,{\overline S _h}^{0{\kern 1pt} 3{\kern 1pt} 3}$\\
\hspace*{80pt}
${\overline S _h}^{2{\kern 1pt} 1{\kern 1pt} 2} =  - \,{\overline S _h}^{1{\kern 1pt} 2{\kern 1pt} 2} = {\overline S _h}^{3{\kern 1pt} 1{\kern 1pt} 3} =  - \,{\overline S _h}^{1{\kern 1pt} 3{\kern 1pt} 3}$\\
where the subscript $h$ says that the components are taken with respect to local orthonormal frames. These relations are also valid for any tensor of rank 3 anti-symmetric with respect to the two first indices. Rotations around the Oy axis can be obtained from the product of the two previous rotations, and will therefore be not considered.

\subsection{Invariance with respect to transvections.}

It is sufficient to consider transvections along the Oz axis only, since transvections in other directions can be deduced by rotation. In hyperbolic space, from the relations (A.3), we have :\\
\hspace*{30pt}
$s{h^2}({\chi _q}) = {(\,sh(\chi )\,ch(L) + ch(\chi )\,sh(L)\,{\mathop{\rm c}\nolimits} \,)^2} + s{h^2}(L)\,{s^2}$\\
\hspace*{30pt}
$sh({\chi _q})\,d{\chi _q} = (\,ch(L)\,sh(\chi ) + c\,sh(L)\,ch(\chi )\,)\,d\chi  + sh(L)\,sh(\chi )\,dc$\\
\hspace*{30pt}
$sh({\chi _q})\,d{c_q} =  - \,\,\frac{{sh(L)\,sh(\chi )\,{s^2}}}{{s{h^2}(\chi_q)}}\,d\chi  + \frac{{s{h^2}(\chi )}}{{s{h^2}(\chi_q)}}\,{(s{h^2}({\chi _q}) - {s^2}\,s{h^2}(L)\,)^{{\textstyle{1 \over 2}}}}\,\,dc$\\
from which we get :\\
\hspace*{30pt}
$\frac{{\partial \,{\chi _q}}}{{\partial \,\chi }} = \frac{{(\,sh(\chi )\,ch(L) + ch(\chi )\,sh(L)\,{\mathop{\rm c}\nolimits} \,)}}{{sh({\chi _q})}} = \sqrt {1 - \frac{{s{h^2}(L)\,{s^2}}}{{s{h^2}({\chi _q})}}} $\\
\hspace*{70pt}
$\frac{{\partial \,{\theta _q}}}{{\partial \,\theta }} = \frac{{sh(\chi )}}{{sh({\chi _q})}}\;\frac{{\partial \,{\chi _q}}}{{\partial \,\chi }}$\\
\hspace*{70pt}
$\frac{{\partial \,{\chi _q}}}{{\partial \,\theta }} =  - \,s\,\,\frac{{sh(L)\,sh(\chi )}}{{sh({\chi _q})}}$
\hspace{120pt} (A.8) \\
\hspace*{70pt}
$\frac{{\partial \,{\theta _q}}}{{\partial \,\chi }} = \,s\,\,\frac{{sh(L)}}{{s{h^2}({\chi _q})}}$\\
\hspace*{70pt}
$\frac{{\partial {\varphi _q}}}{{\partial \chi }} = \frac{{\partial {\varphi _q}}}{{\partial \theta }} = 0$ \hspace{20pt}
$\frac{{\partial \varphi_q}}{{\partial \varphi }} = 1$ \\

With these transformation relations we find that the only nonzero space-time components of a rank 3 tensor, anti-symmetric in the first two indices : ${T^{\alpha {\kern 1pt} \beta {\kern 1pt} \gamma }} =  - \,{T^{\beta {\kern 1pt} \alpha {\kern 1pt} \gamma }}$ ,  are  of the form :\\
\hspace*{90pt}
${T^{0{\kern 1pt} \mu {\kern 1pt} \nu }} =  - \,{T^{\mu \,0{\kern 1pt} \nu }} = q(\eta )\;{\gamma ^{\mu {\kern 1pt} \nu }}$
\hspace{70pt} (A.9) \\
The same calculations can be done in the spherical case, using the relations (A.6) instead of (A.3), leading to equation (A.9).


\begin{thebibliography}{50}

\bibitem{SF} T.P. Sotiriou, V.Faraoni,  Reviews of Modern Physics  82  (2010)  p.451

\bibitem{CS}  S. Capozziello, A. Stabile, Classical and Quantum Gravity 26, 085019 (2009)

\bibitem{BH}  P. Baeckler , F. Hehl , Class. Quantum Grav. 28 (2011) 215017

\bibitem{Pop}  N.J. Poplawski,  Physics Letters B 694  (2010)  p.181 ,  arXiv:1106.4859v1 [gr-qc] 2011

\bibitem{JPP}  The note is available at : hal.archives-ouvertes.fr/hal-01261519

\bibitem{LL}  L. Landau, E. Lifchitz,  Classical Theory of Fields,  Fourth revised english edition , Butterworth Heinemann ed.

\bibitem{Mathpages} mathpages.com/rr/s7-06/7-06.htm

\bibitem{ABS}  R. Adler, M. Bazin, M. Schiffer,  Introduction to General Relativity, Mc Graw-Hill 1965

\bibitem{Wolf} J.A. Wolf,  Spaces of Constant Curvature,  fifth edition, Publish or Perish, Inc. 1984

\end{thebibliography}
\end{document}